\documentclass[pra,twocolumn,showpacs,superscriptaddress]{revtex4-1}

\usepackage[utf8]{inputenc}
\usepackage[english]{babel}
\usepackage[T1]{fontenc}

\usepackage{graphicx}
\usepackage{amsmath}
\usepackage{amsfonts}
\usepackage{amssymb}
\usepackage[breaklinks=true,colorlinks]{hyperref}

\newcommand{\cf}{cf.\ }

\newcommand{\coloneq}{\mathrel{\mathop:}=}
\newcommand{\dd}{\mathrm{d}}
\newcommand{\Tr}{\operatorname{Tr}}
\newcommand{\comm}[2]{\left[{#1},\,{#2}\right]}
\newcommand{\ew}[1]{\left\langle{#1}\right\rangle}
\newcommand{\ket}[1]{\left|{#1}\right\rangle}
\newcommand{\bra}[1]{\left\langle{#1}\right|}

\newcommand{\DC}{\Delta_\mathrm{C}}
\newcommand{\Df}{\Delta_{\mathrm{eff}}}
\newcommand{\ER}{E_\mathrm{R}}
\newcommand{\kR}{k_\mathrm{R}}
\newcommand{\omrec}{\omega_\mathrm{R}}
\newcommand{\kB}{k_\mathrm{B}}
\newcommand{\Ekin}{E_\mathrm{kin}}

\begin{document}

\title{Multistable particle-field dynamics in cavity-generated optical lattices}

\author{Dominik J. Winterauer}
\affiliation{Institut f{\"u}r Theoretische Physik, Universit{\"a}t Innsbruck, Technikerstra{\ss}e~25, A-6020~Innsbruck, Austria}

\author{Wolfgang Niedenzu}
\affiliation{Institut f{\"u}r Theoretische Physik, Universit{\"a}t Innsbruck, Technikerstra{\ss}e~25, A-6020~Innsbruck, Austria}
\affiliation{Department of Chemical Physics, Weizmann Institute of Science, Rehovot~7610001, Israel} 

\author{Helmut Ritsch}
\email{Helmut.Ritsch@uibk.ac.at}
\affiliation{Institut f{\"u}r Theoretische Physik, Universit{\"a}t Innsbruck, Technikerstra{\ss}e~25, A-6020~Innsbruck, Austria}

\begin{abstract}
Polarizable particles trapped in a resonator-sustained optical-lattice potential generate strong position-dependent backaction on the intracavity field. In the quantum regime, particles in different energy bands are connected to different intracavity light intensities and optical-lattice depths. This generates a highly nonlinear coupled particle-field dynamics. For a given pump strength and detuning, a factorizing mean-field approach predicts several self-consistent stationary solutions of strongly distinct photon numbers and motional states. Quantum Monte Carlo wave-function simulations of the master equation confirm these predictions and reveal complex multi-modal photon-number and particle-momentum distributions. Using larger nanoparticles in such a setup thus constitutes a well-controllable playground to study nonlinear quantum dynamics and the buildup of macroscopic quantum superpositions at the quantum-classical boundary.
\end{abstract}

\date{May 19, 2015}

\maketitle

\section{Introduction}

The dynamics of polarizable point-like particles trapped by an optical cavity light field has been the subject of intense theoretical and experimental studies in the past decade~\cite{domokos2003mechanical,ritsch2013cold,aspelmeyer2014cavity,aspelmeyer2014cavitybook}. Beyond implementing improved neutral-atom cavity-QED systems~\cite{ye1999trapping,kruse2003cold,pinkse2000trapping}, recently proposed applications of such setups range from ultrahigh-$Q$ optomechanics~\cite{chang2012ultrahigh,ni2012enhancement,aspelmeyer2014cavity} to precision tests of quantum mechanics at a mesoscopic scale~\cite{romero2011large} and gravity~\cite{pikovski2012probing}. Following the first pioneering experiments more than a decade ago~\cite{ye1999trapping,pinkse2000trapping,kruse2003cold}, several groups have implemented reliable cavity-based optical traps in their experiments for various particle numbers ranging from a single or few atoms~\cite{schleier2011optomechanical,stamper2014cavity,brahms2012optical,thompson2013coupling} to Bose-Einstein condensates (BECs)~\cite{brennecke2007cavity,colombe2007strong,wolke2012cavity,bux2011cavity} or lately even considerably heavier nanoparticles~\cite{asenbaum2013cavity,kiesel2013cavity,millen2015cavity}.

\par

Particles in cavity fields, in contrast to free-space optical potentials, substantially act back on the field dynamics~\cite{domokos2001semiclassical,mekhov2012quantum}, which generates complex and rich nonlinear dynamics~\cite{gupta2007cavity,griesser2011nonlinear,diver2014nonlinear,goldwin2014backaction}. In the standard optomechanical limit of very tightly trapped particles or membranes, which can essentially be modeled by harmonic oscillators~\cite{marquardt2009optomechanics,schulze2010optomechanical}, a wealth of interesting physics beyond ground-state cooling appears in the strong-coupling regime. Typical examples are atom-field entanglement, nonlinear oscillations, and multistable behavior~\cite{marquardt2006dynamical,fernandez2007nonlinear,vukics2009cavity,griesser2011nonlinear,niedenzu2012quantum,dombi2013optical}. The system dynamics gets even more complex and rich, if one refrains from linearizing the particle motion and considers its full dynamics along the cavity potential~\cite{maschler2005cold,niedenzu2010microscopic}. 

\par

In most cases the optical potential along the cavity axis is well approximated by a sinusoidal lattice potential with a depth proportional to the momentary intracavity photon number~\cite{ritsch2013cold}. While for deep potentials the harmonic-oscillator basis allows for analytic insight, it becomes inadequate for shallower lattices. The eigenfunctions of periodic potentials are delocalized Bloch functions, which can be transformed to localized Wannier functions~\cite{kohn1959analytic}. Unfortunately, no analytic solutions for neither the Bloch nor the Wannier functions are known even for a fixed lattice depth. Hence, aiming for an explicit analytic treatment in the (dynamic) quantum-potential limit is a hopeless goal. In view of these complications, several semiclassical and mean-field models with factorized evolution of the particles and the field have been developed to obtain some first insights~\cite{domokos2001semiclassical,griesser2011nonlinear,schuetz2013cooling}. Here the field expectation value is governed by ordinary differential equations containing particle expectation values. This field is in turn inserted in the effective Hamiltonian for the particle motion~\cite{maschler2004quantum,maschler2005cold}. Even in this strongly simplified limit the nonlinearity of the interaction does not allow for a straightforward solution in the general case and further assumptions are needed~\cite{vukics2009cavity}.

\par
 
In this paper we study the full quantum dynamics and the steady-state properties for the case of a single particle in a cavity-sustained optical lattice in the strongly coupled and strongly pumped limit. Hence, our treatment will centrally be based on straightforward numerical solutions of the corresponding quantum-optical master equation. Strong emphasis will be put on steady-state properties of the system in the limit of very low temperatures close to $T=0$, where semiclassical treatments predict a multitude of stationary solutions. To this end we will heavily rely on quantum Monte Carlo wave-function simulations~\cite{dalibard1992wave,dum1992monte,moelmer1993monte}, since a direct solution of the master equation becomes very slow and cumbersome owing to the large joint particle-field Hilbert space, even though we consider the simplest possible system involving only a single particle. 

\par

This paper is organized as follows. In Sec.~\ref{sec_model} we introduce the model Hamiltonian and the master equation, from which we derive equations for the expectation values of the cavity field and the photon number, depending on the particle state. To get some first qualitative insight into the system behavior and to identify interesting parameter regimes, we start with simplified semiclassical models. Factorizing atom and field dynamics, we approximate the photon field by a classical field characterized by its mean photon number, which is determined by the spatial distribution of the particle. We then look for self-consistent steady-state solutions for the expected photon number. Section~\ref{sec_self-consistent} is devoted to studies of these self-consistency conditions in various limiting cases. We first consider the deep-trap limit of harmonic particle confinement, which allows for an analytic treatment. This analysis is afterwards extended to localized Wannier states in the general case of a periodic optical lattice. In Sec.~\ref{sec_numerical} we then numerically solve the full master equation in typical operating regimes determined before and also analyze the behavior of single Monte Carlo trajectories. Finally, in Sec.~\ref{sec_conclusions} the conclusions are drawn.

\section{Model}\label{sec_model}

\par
\begin{figure}
  \centering
  \includegraphics[width=\columnwidth]{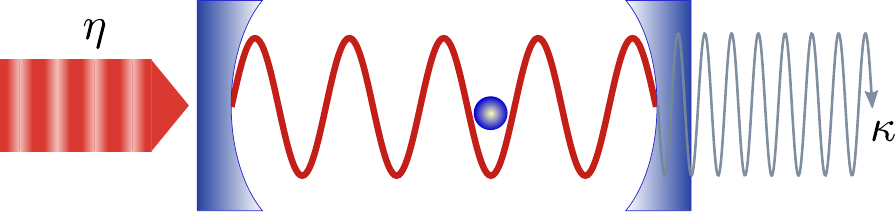}
  \caption{(Color online) A particle within a driven optical cavity. The longitudinal cavity pump $\eta$ builds up an intracavity field that drives the particle motion. The particle's motional state affects the cavity detuning (dynamic refractive index), which in turn influences the intracavity photon number. Photons leak through the cavity mirrors at a rate $2\kappa$.}
  \label{fig_system}
\end{figure}
\par

We consider the standard case of a driven, damped cavity mode with a single polarizable particle of mass $m$ moving along the cavity axis, as sketched in Fig.~\ref{fig_system}. The Hamiltonian ($\hbar=1$) in a rotating frame with pump amplitude $\eta$, cavity detuning $\DC$, and effective particle-field interaction strength $U_0$ is then given by~\cite{hechenblaikner1998cooling}
\begin{equation}
  \label{Hamiltonian}
  H = \frac{p^2}{2m}-\left[\DC-U_0 \cos^2(\kR x)\right]a^{\dagger}a-i\eta\left(a-a^{\dagger}\right),
\end{equation}
where $\kR=2\pi/\lambda$ is the single-photon recoil momentum, with $\lambda$ being the cavity mode wavelength. The particle position and momentum operators $x$ and $p$, and the photon mode annihilation (creation) operators $a^{\left(\dagger\right)}$, obey the standard canonical commutation relations
\begin{equation}
  \comm{x}{p}=i , \;
  \comm{a} {a^{\dagger} }=1,
\end{equation}
with all other commutators vanishing. The coupling strength is parametrized by $U_0$, which denotes the optical potential depth per photon as well as the maximum cavity mode resonance frequency shift that a particle induces when placed at a field antinode. Here we consider a large negative $U_0$, which corresponds to high-field seeking particles.

\par

We assume operation far from any internal optical resonances, such that spontaneous light scattering and absorption losses from the particle into the mode can be neglected~\cite{domokos2003mechanical}. The dominant loss mechanism is then cavity damping, which at optical frequencies can be incorporated by a standard master equation treatment parametrized by a loss rate $\kappa$~\cite{gardinerbook},
\begin{equation}\label{eq_master}
  \dot{\rho} = -i \comm{H}{\rho}+\kappa \left(2a \rho a^{\dagger}-a^{\dagger}a\rho-\rho a^{\dagger}a \right).
\end{equation}

\par

From this master equation we straightforwardly derive ordinary differential equations (ODEs) for the expectation values of the field amplitude, $\ew{a}=\alpha$, and the photon number, $\ew{n}=\ew{a^{\dagger} a}$, which read
\begin{subequations}\label{eq_alpha_n}
  \begin{align}
    \dot{\alpha}&=\left[i\left(\DC-U_0\ew{\cos^2(\kR x)}\right)-\kappa\right]\alpha+\eta \\
    \dot{\ew{n}}&=\eta\left(\alpha+\alpha^{\ast}\right)-2\kappa\ew{n}.
  \end{align}
\end{subequations}
Within the semiclassical treatment with a $c$-number description of the field amplitude the particle-field density matrix is assumed to be separable. Obviously, the field dynamics depends on the motional state of the particle via the expectation value (bunching parameter)
\begin{equation}\label{eq_ew_cos2}
  b\coloneq\ew{\cos^2(\kR x)}
\end{equation}
in a nonlinear fashion. This parameter itself is, in turn, governed by the Schr{\"o}dinger equation containing the Hamiltonian~\eqref{Hamiltonian}, whose spatial eigenstates are Bloch functions according to the quantized lattice depth $V_0= |U_0| a^{\dagger}a$. This yields a different evolution for each photon-number component of the total wave function and thus a very complex time evolution. Hence, a full solution of the master equation~\eqref{eq_master} requires a numerical approach, which can be directly implemented using a truncated photon number and momentum basis expansion. Note that due to the periodic nature of the potential we can work with periodic boundary conditions in real space and use a discrete momentum basis $\{\ket{p}=\ket{j\kR}\}$ with $j \in \mathbb{N}_0$.

\par

As these calculations are time consuming and the range of physical parameters $(\eta,\DC,U_0,\omrec)$ is large, we first try to get some qualitative insight and find interesting parameters regions using the factorized semiclassical approach involving Eqs.~\eqref{eq_alpha_n}.

\section{Self-consistent semiclassical solutions of the coupled atom-field dynamics}\label{sec_self-consistent}

Let us now analyze potential stationary solutions of the coupled ODE system~\eqref{eq_alpha_n}. As the field dynamics in the semiclassical approximation depends on the position distribution of the particle via the expectation value~\eqref{eq_ew_cos2} only, for the system to reach a steady state we need a stationary wave function. This leads to the self-consistency condition
\begin{equation}
  \ew{n} = \frac{\eta^2}{\kappa^2+\left(\DC-U_0\ew{\cos^2(\kR x)}\right)^2},
\label{eq_selcons}
\end{equation}
where the wave function of the particle has to be an eigenstate of Eq.~\eqref{Hamiltonian} with the photon-number operator $a^{\dagger}a$ replaced by $\langle n\rangle$. Note that the expectation value $\ew{\cos^2(\kR x)}$ in the denominator on the right-hand side of Eq.~\eqref{eq_selcons} does not explicitly involve any field operators. Nevertheless, the time evolution of the spatial part of the wave function depends on the field intensity. Hence, the state can only be stationary if it is an eigenstate of the Hamiltonian~\eqref{Hamiltonian} for the momentary photon number. Note that the pump amplitude $\eta$ is a free parameter in the above equation and in many cases for a given eigenstate of the particle Hamiltonian a self-consistent choice of $\eta$ can be made to fulfill Eq.~\eqref{eq_selcons}~\cite{maschler2004quantum}. We, however, opt for the opposite approach and determine self-consistent photon numbers for given pump strengths. 

\par

Let us mention, though, that this is only a necessary condition and by no means sufficient for a stable stationary equilibrium subject to the quantum fluctuations of the system. At this point it can only serve as a guide towards interesting parameter regions, which is, e.g., the case when several different spatial eigenfunctions lead to the same pump amplitude $\eta$. We will discuss this in some more detail below for specific limiting examples. 

\subsection{Harmonic-oscillator expansion in a deep lattice}

\begin{figure}
  \centering
  \includegraphics[width=\columnwidth]{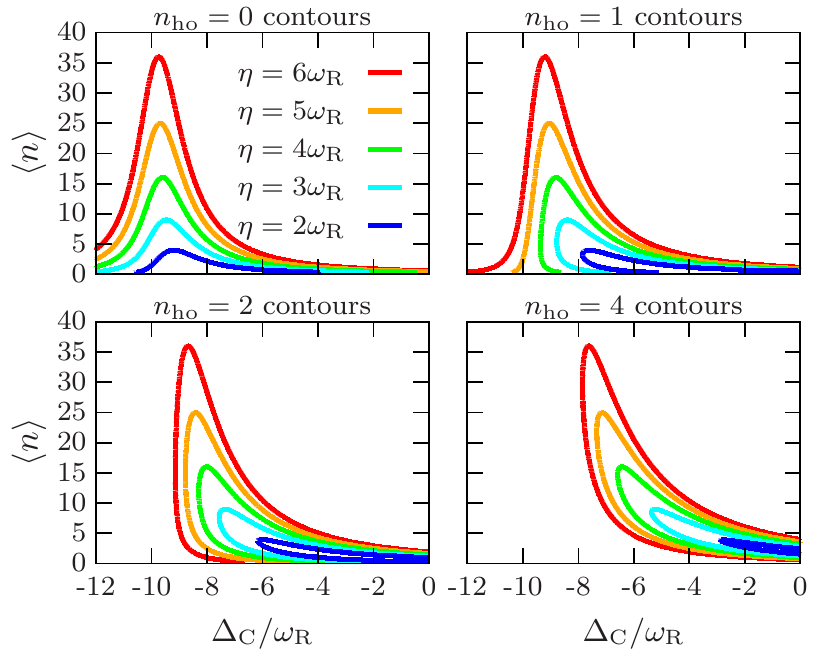}
  \caption{(Color online) Self-consistent photon-number contours for a harmonically trapped particle as a function of the cavity-pump detuning for different pump amplitudes $\eta$. The four plots show contours in the $\DC$-$\ew{n}$ plane for different harmonic-oscillator eigenstates $\ket{n_{\text{ho}}}$, where Eq.~\eqref{eq_selconsho} holds self-consistently. All excited states, $n_{\text{ho}}\geq 1$, yield the possibility of two self-consistent solutions for a certain range of detuning $\DC$ and therefore may allow for optomechanical bistability. Parameters: $U_0=-10\omrec$ and $\kappa=\omrec$.}
  \label{fig_selconsho}
\end{figure}

In the limit where the potential depth $V_0 \approx |U_0| \ew{n} $ strongly exceeds the recoil energy $\ER\equiv\omrec\coloneq \kR^2 /(2 m)$, the lowest-energy particle states are well localized within a single well of the optical lattice. For low enough temperatures the optical potential $V_{\text{eff}}(x)=U_0\ew{n}\cos^2(\kR x)$ can then be approximated by a harmonic potential~\cite{maschler2004quantum},
\begin{equation}
  U_0\ew{n}\cos^2(\kR x) \approx U_0\ew{n}\left(1-\kR^2 x^2\right).
\end{equation} 
The corresponding trapping frequency $\omega_{\mathrm{ho}}$ then reads
\begin{equation}
  \frac{\omega_{\mathrm{ho}}}{\omrec} = 2\sqrt{\frac{|U_0|}{\omrec}\ew{n}} \gg 1,
\end{equation}
and we can analytically find the respective oscillator states $\ket{n_{\mathrm{ho}}}$ to this frequency. The expectation value in the denominator of Eq.~\eqref{eq_selcons} is then well approximated by $\ew{\cos(\kR x)^2} \approx 1 - \kR^2 \ew{x^2}$ with 
\begin{equation}
  \ew{x^2}_{n_{\mathrm{ho}}}=\bra{n_{\mathrm{ho}}}x^2\ket{n_{\mathrm{ho}}}=\frac{2n_{\mathrm{ho}}+1}{2m\omega_{\mathrm{ho}}},
\end{equation}
such that
\begin{equation}
  \kR^2\ew{x^2}_{n_{\mathrm{ho}}}=\left(2n_{\mathrm{ho}}+1\right)\frac{\omrec}{\omega_{\mathrm{ho}}}=\frac{2n_{\mathrm{ho}}+1}{2\sqrt{\frac{|U_0|}{\omrec}\ew{n}}}.
\end{equation}
Hence, within the harmonic-oscillator approximation Eq.~\eqref{eq_selcons} becomes the simple algebraic equation
\begin{align}
  \ew{n} = \frac{\eta^2}{\kappa^2+\left[\DC-U_0\left(1-\frac{2n_{\mathrm{ho}}+1}{2\sqrt{\frac{|U_0|}{\omrec}\ew{n}}}\right)\right]^2},
\label{eq_selconsho}
\end{align}
which can be easily solved for each choice of eigenstate number $n_\mathrm{ho}$. Figure~\ref{fig_selconsho} shows contours in the $\DC$-$\ew{n}$ plane for different values of $\eta$ for which Eq.~\eqref{eq_selconsho} holds. While the lowest-energy state $n_\mathrm{ho}=0$ results in a unique photon number, multiple (up to two) solutions are possible for higher excited states.

\par

\subsection{Self-consistent states for the full lattice dynamics}

Several interesting aspects of the deep-trap harmonic-oscillator regime have been studied in the past~\cite{maschler2004quantum,diver2014nonlinear,vukics2009cavity}. In many respects the system is directly related to standard optomechanical models with quadratic coupling~\cite{marquardt2009optomechanics}. For ultracold particles and weaker optical potentials the motion is strongly delocalized in the lattice~\cite{ritsch2013cold}. Hence, we now turn to the full model and consider the particle's motion in the periodic optical lattice
\begin{equation}
  V_{\text{eff}}(x) = U_0\ew{n}\cos^2(\kR x).
\end{equation}

\par

For very shallow lattices close to zero temperature, i.e., for a BEC in a cavity, a two-mode expansion of the wave function can be applied~\cite{brennecke2007cavity,szirmai2010quantum}, which again allows for analytic treatments and analogies with optomechanical couplings. However, the validity range of this model is limited in temperature, time, and coupling strength. As we are here more interested in the limit of strong nonlinear backaction in deep potentials, we cannot apply this simplification and have to solve the Schr{\"o}dinger equation for a periodic potential, which gives us the well-known Bloch states $\Psi_{mq}(x)$, where $m$ denotes the energy band and $q$ is the quasi-momentum~\cite{kohn1959analytic}. Being periodic with the lattice constant, they are not the best basis to describe a single localized particle. Hence, we switch to a Wannier basis, where each basis state represents a localized wave function with its center of mass at a particular lattice site. Such basis states have been very successfully used to study ultracold particle dynamics in optical lattices~\cite{jaksch1998cold,zwerger2003mott}. The Wannier functions for a given band index $m$ localized at lattice position $R$ are defined as~\cite{kohn1959analytic}
\begin{equation}
  w_m(x-R):=\sqrt{\frac{a}{2\pi}}\int_{-\pi/a}^{\pi/a}\Psi_{mq}(x)e^{-iqR}\dd q,
\end{equation}
where $a$ is the lattice periodicity. The Bloch functions $\Psi_{mq}(x)$ are only defined up to a phase. In order to obtain the \emph{maximally localized} (i.e., real and exponentially decaying) Wannier functions, these phases need to be properly adjusted~\cite{kohn1959analytic}. In what follows we choose for simplicity $R=0$.

\par

\begin{figure}
  \centering
  \includegraphics[width=\columnwidth]{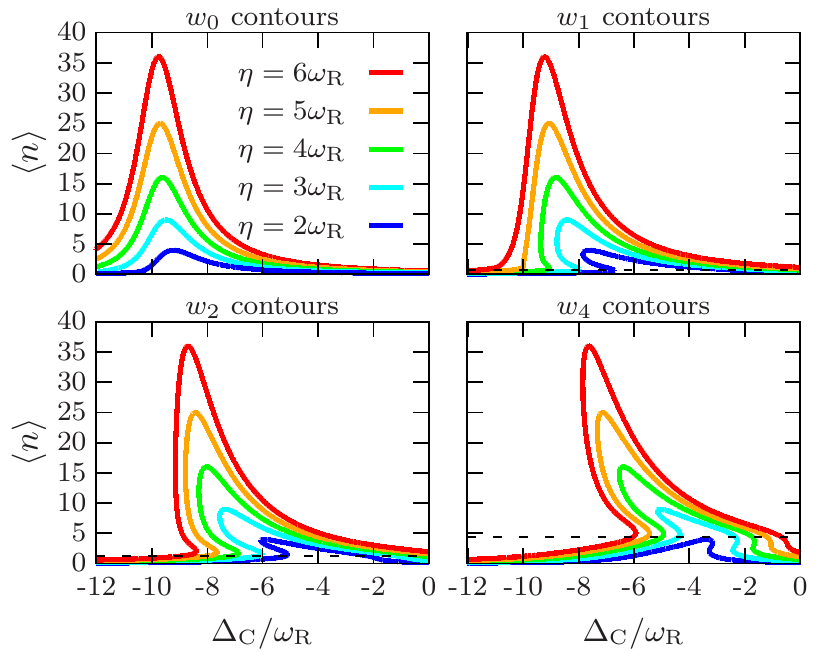}
  \caption{(Color online) Same as Fig.~\ref{fig_selconsho} for particles in localized Wannier states. Photon numbers above the dashed lines for $w_{m \geq 1}$ correspond to bound Wannier states ($E<0$). Higher bands exhibit up to three solutions of Eq.~\eqref{eq_selconswa2} for a given value of $\DC$. Parameters: $U_0=-10\omrec$ and $\kappa=\omrec$.}
  \label{fig_selconswa}
\end{figure}

We are now able to restate the self-consistency equation~\eqref{eq_selcons} for each band index $m$ as
\begin{equation}
  0=\frac{\eta^2}{\kappa^2+\bigl(\DC-U_0b_m\bigr)^2}-\ew{n},
  \label{eq_selconswa2}
\end{equation}
with
\begin{equation}
  b_m = \int_{-\infty}^{\infty}[w_m(x)]^2\cos^2(\kR x)\,\dd x.
  \label{eq_potexp}
\end{equation}
Contrary to the harmonic oscillator wave functions, there is no analytic expression for Wannier functions and we have to numerically solve the Schr{\"o}dinger equation for each particular $\ew{n}$. Therefore $\ew{n}$ does not explicitly appear on the right-hand side of Eq.~\eqref{eq_selconswa2}, but enters implicitly through the shape of the wave function. As before, we can obtain the contours where Eq.~\eqref{eq_selconswa2} holds self-consistently in the $\DC$-$\ew{n}$ plane for the same values of $\eta$; see Fig.~\ref{fig_selconswa}.

\par

\par

\begin{figure}
  \centering
  \includegraphics[width=\columnwidth]{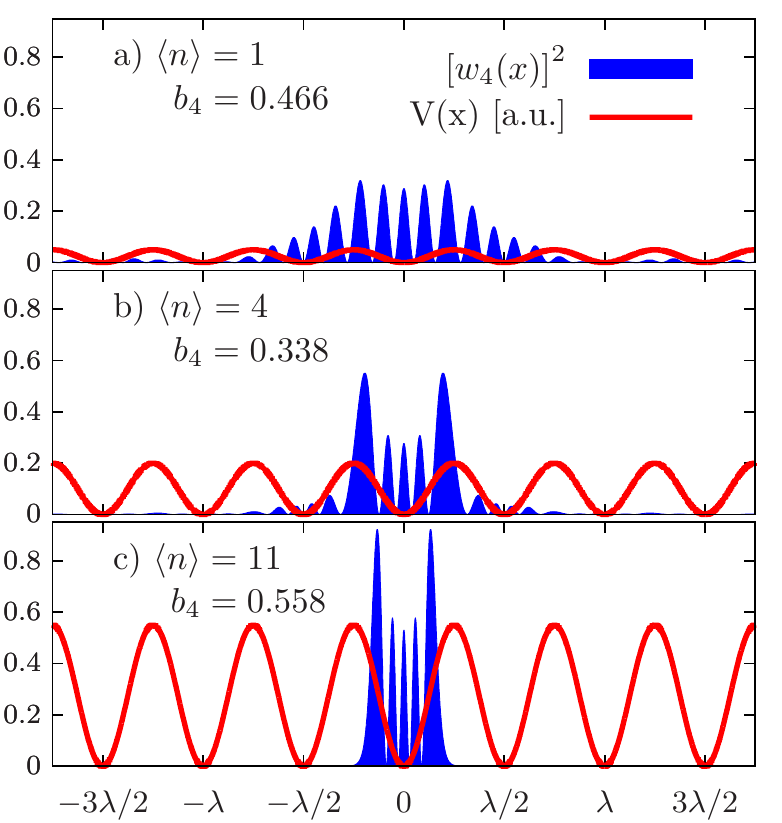}
  \caption{(Color online) Spatial particle distribution in a fourth band Wannier state for different photon numbers: (a) Free particle: The wave function is barely localized, $b_4 \lesssim 0.5$. (b) Transition to a bound state: The wave function is localized at potential \emph{maxima}, $b_4 < 0.5$. (c) Tight-binding regime: The wave function is strongly localized in a single potential well, $b_4 > 0.5$.}
\label{fig_wannloc}
\end{figure}

The behaviors of the photon numbers for the lowest-energy states $n_{\mathrm{ho}}=0$ and $m=0$ in both cases are very much alike, because the corresponding lowest bound states in both models are similar. Indeed, the maximally localized Wannier functions converge to the harmonic oscillator functions for deep potentials~\cite{kohn1959analytic}. For the higher-energy eigenstates, however, the photon numbers differ significantly. The reason behind is that in a harmonic oscillator all states are bound, while Wannier states for increasing $m \geq 1$ undergo a transition from bound to free states for a given photon number (i.e., potential depth). Dashed lines in Fig.~\ref{fig_selconswa} indicate this boundary. For $m \geq 1$ sharp bends appear, yielding self-consistent contours reminiscent of nonlinear response curves. The origin of these peculiarity at the transition from free to bound states becomes evident, if we look at the spatial particle density of the respective Wannier states. Figure~\ref{fig_wannloc} illustrates the behavior of the fourth band Wannier state $w_4$ for different mean photon numbers (i.e., potential depths). The key quantity here is the expectation value of the bunching parameter $b_m$ [Eq.~\eqref{eq_potexp}], which determines the backaction of the particle on the cavity field, i.e., its effective refractive index. For free particles, $\ew{E}>0$, the wave function is barely localized and $b_m \approx \frac{1}{2}$. Around $\ew{E}\approx 0$ the Wannier states localize around potential maxima, i.e., optical field nodes, which minimizes the backaction of the particle on the cavity, $b_m<\frac{1}{2}$, while for deeper potentials $\ew{E}<0$ and particles are drawn towards field antinodes and the index of refraction increases with potential depth, $b_m>\frac{1}{2}$. Thus the nonlinear behavior of the refractive index allows for multiple self-consistent solutions for certain ranges of the cavity detuning $\DC$. In particular, we also find solutions corresponding to unbound particle states (e.g., for $w_4$ in Fig.~\ref{fig_selconswa}).

\par

\subsection{Stability of the self-consistent factorized solutions}

\begin{figure}
  \centering
  \includegraphics[width=\columnwidth]{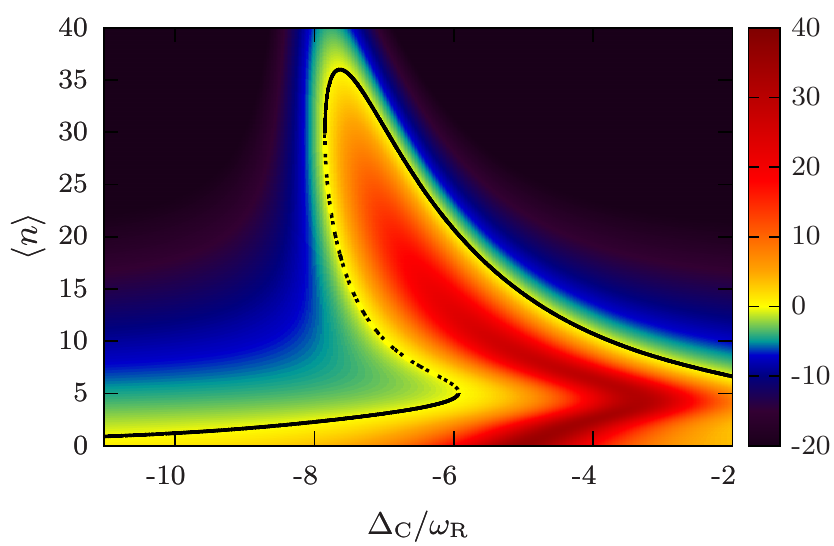}
  \caption{(Color online) Contour plot for the right-hand side of Eq.~\eqref{eq_selconswa2} for the fourth band Wannier state. Solid (dotted) lines mark stable (unstable) self-consistent solutions according to Eq.~\eqref{eq_selconstab}. Parameters: $\eta = 6\omrec$, $U_0=-10\omrec$, and $\kappa=\omrec$.}
  \label{fig_selcons3d}
\end{figure}

As we saw above, for certain parameter ranges in both the harmonic oscillator and the Wannier contour plots more than one self-consistent solution appears. Whether or not these solutions have significant relevance for the full system dynamics depends on their stability properties, i.e., their response against small deviations in the photon number or the spatial distribution. Some qualitative insight can already be gained by virtue of Eq.~\eqref{eq_selcons}. The right-hand side of Eq.~\eqref{eq_selcons} depends on the shape of the wave function in real space, which in our semiclassical model implicitly depends on $\ew{n}$. The term on the right-hand side of Eq.~\eqref{eq_selcons} determines the mean photon number that is allowed by the spatial part of the wave function in steady state. If it increases or decreases with $\ew{n}$ faster than $\ew{n}$ at a self-consistent point, one may assume that the self-consistent configuration is unstable. Therefore we find that at \textit{stable} self-consistent configurations the inequality
\begin{align}
\frac{\partial}{\partial \ew{n}} \frac{\eta^2}{\kappa^2+\left(\DC-U_0\ew{\cos^2(\kR x)}\right)^2} -1 < 0
\label{eq_selconstab}
\end{align}
must hold. This rather intuitive result is verified in Appendix~\ref{app_stability} via linear stability analysis. The stability regions for the fourth band (where up to three self-consistent solutions exist) are shown in Fig.~\ref{fig_selcons3d}.

\par

\section{Numerical analysis of the full coupled atom-field dynamics}\label{sec_numerical}

In order to test the above analysis, we now strive to solve the full master equation [Eq.~\eqref{eq_master}]. As already mentioned, the (even for a single particle) large Hilbert space makes a direct numerical integration attempt practically unfeasible for realistic parameters and photon numbers. We therefore make use of quantum Monte Carlo wave-function simulations, in which single stochastic state vectors (instead of the whole density matrix) are evolved subject to a non-Hermitian effective Hamilton operator~\cite{dalibard1992wave,dum1992monte,moelmer1993monte}. This evolution is stochastically interrupted by quantum jumps corresponding to a projective removal of one photon. Mathematically, averaging over a large number of such trajectories then approximates the full density matrix. Interestingly, the jumps can be physically interpreted as detection events of photons leaking out of the resonator. Hence, single trajectories provide a microscopic view of the processes incorporated in the master equation since the ensemble average over many trajectories converges towards the solution of the latter.

\par

In what follows we compare our above predictions with the time evolution of single trajectories as well as to their ensemble average. The numerical implementation of these simulations was done within the C++QED framework allowing efficient and fast simulations~\cite{vukics2007cppqed,vukics2012cppqedv2,sandner2014cppqedv2}. Since dynamic aspects have been eliminated in our self-consistency and stability analysis, it is not clear which of the self-consistent solutions appear in the dynamics and what are their corresponding probabilities.

\subsection{Time evolution of single trajectories}

\begin{figure}
  \centering
  \includegraphics[width=\columnwidth]{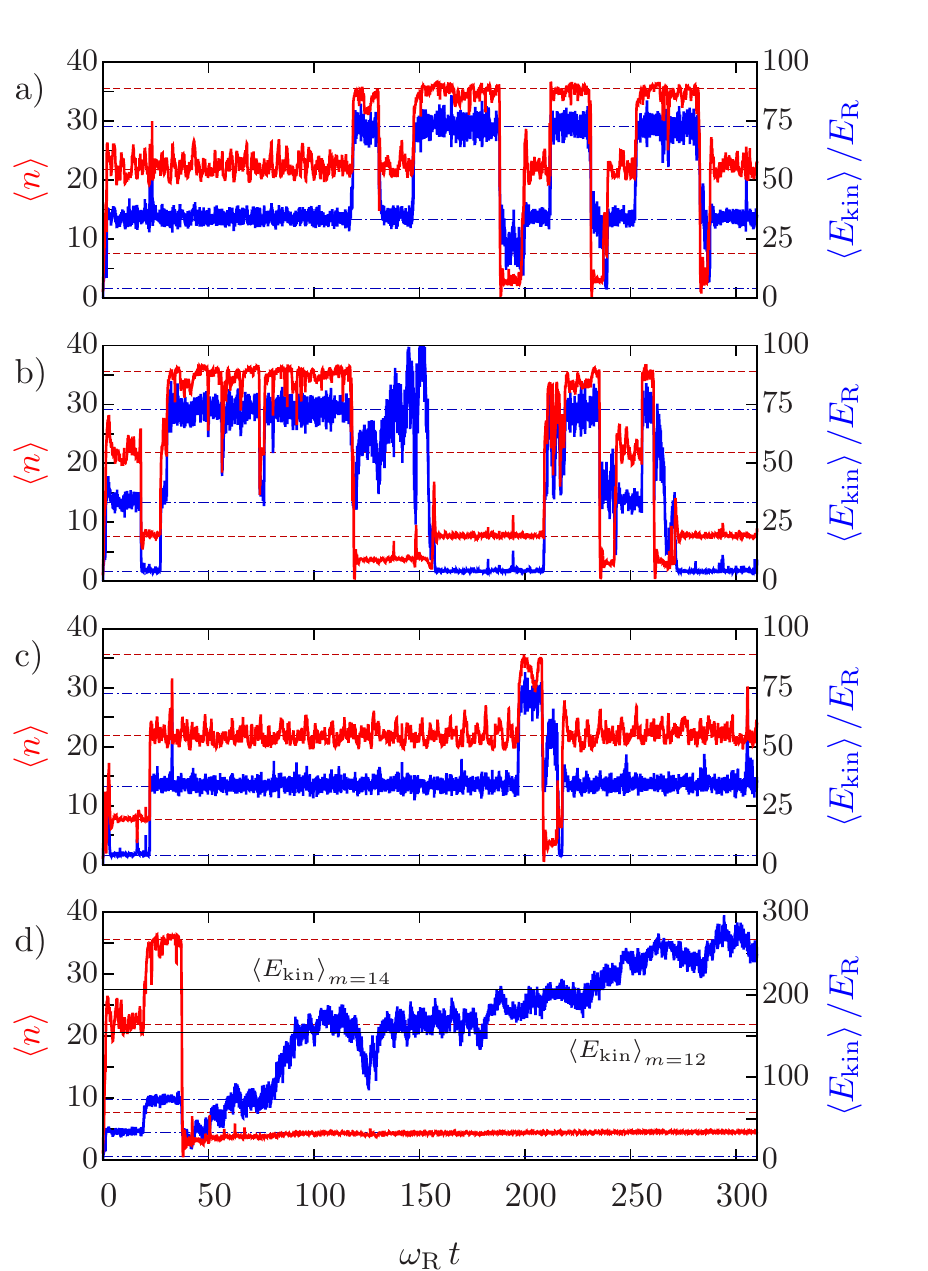}
  \caption{(Color online) Expectation values on single quantum trajectories for $\eta=6\omrec$, $\DC=-7.5\omrec$, $U_0=-10\omrec$, and $\kappa = \omrec$. The mean values jump between the stable values predicted by the self-consistency equation~\eqref{eq_selconswa2} (dashed lines). (a) and (b) Trajectories with several jumps from very high to very low photon numbers. (c) Trajectory with only a few jumps. (d) Trajectory with a large increase of $\ew{E_{\mathrm{kin}}}$ during evolution in a low-photon-number state. The kinetic energy mean value of the 12th and 14th excited bands are shown as black solid lines. Although numerical values indicate that the trajectory could be in a definite band, the neat picture of correlated photon number and momentum jumps clearly breaks down at this point. Note the different scaling of the $\ew{E_{\mathrm{kin}}}$ axis.}
  \label{fig_singtraj}
\end{figure}

We consider a small sample of single quantum trajectories in a multistable regime. Figure~\ref{fig_singtraj} shows the corresponding expectation values of the intracavity photon number $\ew{n}$ as well as of the kinetic energy $\ew{E_{\mathrm{kin}}}=\ew{p^2}/(2m)$. As one might expect, both quantities jump simultaneously between rather stable values. The latter can be identified as the possible semiclassical values found above. Each trajectory thus seems to switch between these states rather than forming state superpositions. Between jumps both quantities appear to fluctuate only weakly about the self-consistent values (upper three graphs). In some cases $\ew{n}$ jumps to very low values, where no bound state exists. In such cases the system continuously heats up (i.e., $\ew{E_{\mathrm{kin}}}$ increases) until a subsequent jump occurs and projects the particle back into a bound state (as for example in Fig.~\ref{fig_singtraj}b between the two quantum jumps at $\omrec t\approx 120$ and $150$; the significantly increased photon number after the second jump allows again for bound states). Figure~\ref{fig_singtraj}d shows an extreme case, where the particle remains essentially free for a long time. We find that there exists a multitude of stable self-consistent solutions of Eq.~\eqref{eq_selconswa2} around $\ew{n}=4$ for higher bands ($m \geq 6$), whose self-consistent photon numbers increase only slightly with increasing band index. A small plateau of $\ew{E_{\mathrm{kin}}}$ in Fig.~\ref{fig_singtraj}d can be interpreted as an occupation of the 12th Wannier state, $\ew{E_{\mathrm{kin}}}_{m=12}$. We also indicate the mean kinetic energy of the 14th excited band, $\ew{E_{\mathrm{kin}}}_{m=14}$, and observe that this energy is reached continuously rather than by discrete jumps as in the bound case. Below we will re-encounter reminiscences of such trajectories in the ensemble-averaged solution of the master equation.

\par

Due to the self-consistent photon numbers' small sensitivity on the band index for $m \geq 6$, according to our semiclassical analysis several excited bands can co-exist at a given value of $\ew{n}$. Wannier functions for a given potential depth are mutually orthogonal, therefore transitions between bands can only occur through fluctuations in the unbound regime, yielding much slower transition rates than in the bound regime. Trajectories may jump back to bound states, Figure~\ref{fig_singtraj}a--c, or remain unbound, Figure~\ref{fig_singtraj}d. The likeliness of jumping back to a bound state seems to decrease with kinetic energy. At this stage it appears that the momentum part of the wave function controls the expected intracavity photon number rather than vice versa.

\par

The correlated particle-field jumps reflect strong particle-field correlations and some amount of entanglement, as previously discussed in similar contexts~\cite{vukics2007microscopic,niedenzu2012quantum}.

\par

\subsection{Stationary solution of the master equation via ensemble-averaged quantum trajectories}

\begin{figure}
  \centering
  \includegraphics[width=\columnwidth]{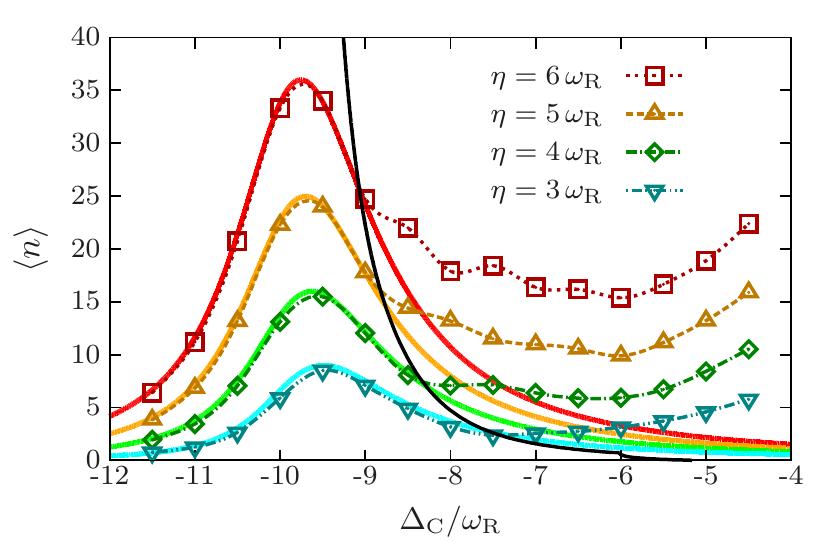}
  \caption{(Color online) Self-consistent solutions of Eq.~\eqref{eq_selconswa2} for the lowest Wannier states ($m=0$) compared to ensemble-averaged quantum simulation results. Solid lines mark self-consistent contours; data points are quantum simulation results. The black line separates the regions where $\Delta_{\mathrm{eff},\,0} < -\Delta_{\mathrm{eff},\,2}$ (left) and $\Delta_{\mathrm{eff},\,0} > -\Delta_{\mathrm{eff},\,2}$ (right) along the self-consistent contours of the zeroth band; \cf Eq.~\eqref{eq_wanndet}. In the left region quantum simulations are in accordance with the self-consistent solutions of Eq.~\eqref{eq_selconswa2} for the lowest band, while deviations arise in the right region due to population of the $m=2$ Wannier state. Parameters: $U_0=-10\omrec$ and $\kappa=\omrec$.}
  \label{fig_selconsnbar}
\end{figure}

We now investigate the solution of the master equation~\eqref{eq_master} of the joint particle-field density matrix by averaging over a sufficiently large ensemble of Monte Carlo trajectories. First we check the distribution of photon numbers for a specific choice of parameters and compare it to the semiclassical results. In Fig.~\ref{fig_selconsnbar} we depict the simplest case of parameters, where only a single semiclassical solution exists. Interestingly, we see that the mean photon numbers obtained from the Monte Carlo simulations agree surprisingly well with the self-consistent solutions of Eq.~\eqref{eq_selconswa2} for the lowest Wannier state, as long as the cooling regime (large negative effective detuning) is maintained. Closer to resonance we see a deviation towards higher photon numbers, which indicates the appearance of motional exited states (\cf Fig.~\ref{fig_selconswa}).

\begin{figure}
  \centering
  \includegraphics[width=\columnwidth]{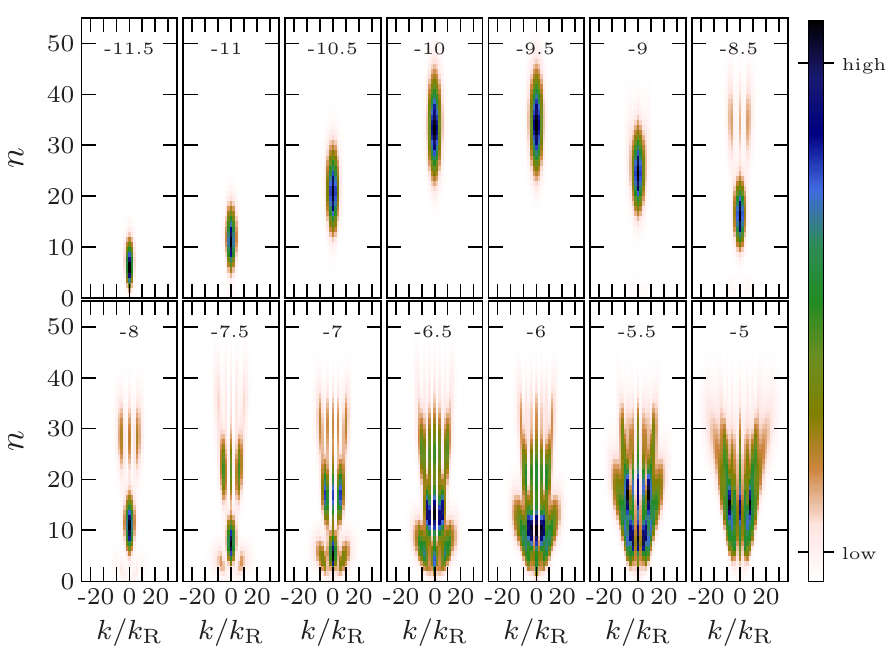}
  \caption{(Color online) Combined photon-number and momentum-state occupation probability after a time interval of $\Delta t= 310\omrec^{-1}$, starting from a state with zero momentum and one cavity photon, $k=0$ and $n=1$, for $U_0=-10\omrec$, $\kappa = \omrec$, $\eta = 6\omrec$, and different values of the detuning. The density matrix is approximated via quantum simulations with ensemble averages over 1000 trajectories for each parameter set. The numbers in each box give the detuning $\DC$ in units of $\omrec$.}
  \label{fig_rhoraywann6}
\end{figure}

\begin{figure}
\centering
\includegraphics[width=\columnwidth]{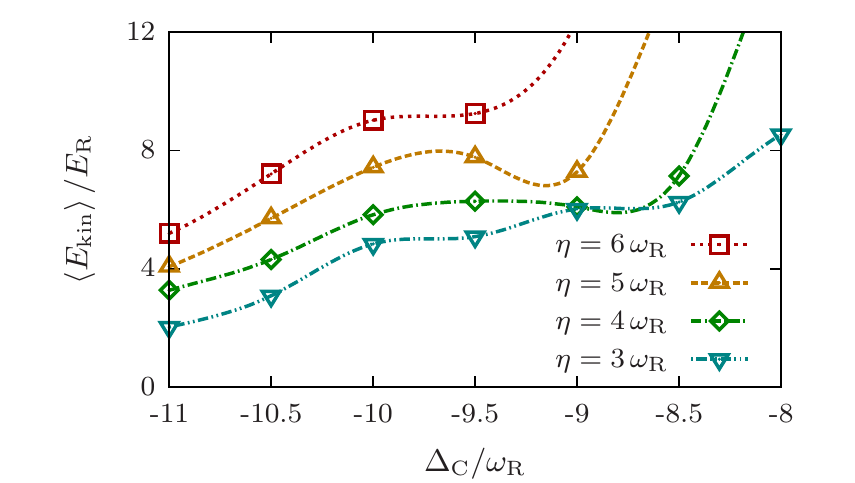}
\caption{(Color online) Quantum simulation results showing the particle's average kinetic energy $\ew{E_{\mathrm{kin}}}$ as a measure of its temperature as a function of the cavity detuning for different pump amplitudes. For $\eta=5\omrec$ an interval where the final temperature decreases with increasing detuning is clearly visible. The plot shows only a detail of the set of data points for better visibility of the effect. In order to eliminate temporal fluctuations at single time instances the plotted points are time averages of $\ew{E_{\mathrm{kin}}}$ of 100 values in the interval $\omrec t\in\left(305,310\right]$. The dashed lines are interpolations between data points and merely serve as a guide to the eye. The other parameters are $U_0=-10\omrec$ and $\kappa=\omrec$. }
\label{fig_cpppvar}
\end{figure}

Motivated by the single trajectories depicted in Fig.~\ref{fig_singtraj} one can deduce a microscopic interpretation of the dynamics. To each band may be assigned a kinetic temperature $\kB T=2\Ekin=\ew{p^2}/m$, which increases with the band index~\footnote{A single particle does not have a temperature; neither does a single-particle quantum state. When we speak of temperature we think of averaging over a fictitious ensemble.}. The sign of the effective detuning
\begin{equation}\label{eq_def_deltaeff}
  \Delta_{\mathrm{eff},\,m} \coloneq \DC-U_0b_m\bigr.
\end{equation}
determines whether the according band is heated ($+$) or cooled ($-$). Heating means that in a certain band the system tends towards populating higher excited bands, while cooling implies the opposite. Since the value of $\Delta_{\text{eff},\,m}$ is different for every band, some bands (the lower ones) are heated and others (the higher ones) are cooled. From the proportionality of the cooling/heating rates to $\Delta_{\text{eff},\,m}$, we conclude that higher bands appear in the ensemble-averaged steady-state solution if
\begin{equation}
  \Delta_{\mathrm{eff},\,m} > -\Delta_{\mathrm{eff},\,m+2}.
  \label{eq_wanndet} 
\end{equation}
Note that for symmetry reasons the dynamics induced by the Hamiltonian~\eqref{Hamiltonian} conserves the parity of the initial state. For a particle initially in the ground state, the lowest accessible excited state is the second band and consequently $m+2$ appears in Eq.~\eqref{eq_wanndet}. Hence the system effectively remains in the lowest band until $\Delta_{\mathrm{eff},\,0} > -\Delta_{\mathrm{eff},\,2}$; see Figs.~\ref{fig_selconsnbar} and~\ref{fig_rhoraywann6}. This implies that, though the system is effectively blue detuned, it does not get heated; see Fig.~\ref{fig_cpppvar}. For certain parameter values a further increase of $\DC$ around $\Delta_{\mathrm{eff},\,0}=0$ even yields further cooling before the second excited band is populated.

\par

\par

\par

A more complete picture of this dynamics can be found if one includes momentum distributions, as depicted in Fig.~\ref{fig_rhoraywann6}, which represents the essence of the underlying physics. For large negative cavity detunings $\DC$ the photon-number distribution follows the mean values obtained for the lowest-band approximation from our semiclassical model [Eq.~\eqref{eq_selconswa2}]. With increasing $\DC$ higher bands $m+2$ appear when heating depletes the lowest band if Eq.~\eqref{eq_wanndet} is satisfied. Note that the population of higher momenta at relatively small photon number in Fig.~\ref{fig_rhoraywann6} (e.g., at $\DC=-7.5\omrec$) can be traced back to trajectories like the one shown in Fig.~\ref{fig_singtraj}d.

\section{Conclusions and outlook}\label{sec_conclusions}

We have shown that the dynamics of a quantum particle trapped in a cavity-sustained optical lattice can reach a multitude of quasi-stationary solutions at the same operation parameters. Fluctuations of the cavity field as well as quantum dynamics of the particle eventually trigger transitions between such states observable in single Monte Carlo quantum trajectories and stable for extended periods. Key properties of these strongly correlated atom-field solutions can be understood from an analysis in terms of localized Wannier functions and a mean-field approximation of the cavity mode. Quantum simulations exhibit few but fast transitions between such quasi-stationary states. Averaged over a sufficiently large ensemble, the final density matrix in this regime is a mixture of several Bloch bands with corresponding photon-number distributions. While the density matrix is mostly a mixture of such quasistationary states, some atom-field entanglement can be present in the transition phase, where the photon-number expectation value lies in between two stationary values.

\begin{acknowledgments}

This work has been supported by the Austrian Science Fund FWF through the SFB F4013 FoQuS.

\end{acknowledgments}

\appendix

\section{Linear stability analysis of self-consistent solutions}\label{app_stability}

We here present the derivation of Eq.~\eqref{eq_selconstab}. The equations of motion for the field amplitudes $\boldsymbol{\alpha}=\left(\alpha,\alpha^{\ast}\right)^T$ read
\begin{equation}
  \frac{\partial}{\partial t} \boldsymbol{\alpha} = U(\alpha,\,\alpha^{\ast})\boldsymbol{\alpha}+\boldsymbol{\eta},
\end{equation}
with $\boldsymbol{\eta}=\left(\eta,\eta\right)^T$ and
\begin{equation}
  U(\alpha,\,\alpha^{\ast})=
  \begin{pmatrix}
    i\Df(\alpha,\,\alpha^{\ast})-\kappa & 0 \\
    0 & -i\Df(\alpha,\,\alpha^{\ast})-\kappa
  \end{pmatrix}.
\end{equation}
Its steady-state solution reads
\begin{equation}\label{eq_app_alpha_ss}
  \boldsymbol{\alpha}_{\mathrm{ss}}=\begin{pmatrix}
    \frac{-\eta}{i\Df-\kappa} \\
    \frac{-\eta}{-i\Df-\kappa}
  \end{pmatrix}.
\end{equation}
We now investigate the stability of this steady-state solution against small perturbations. The effective detuning $\Df\coloneq\DC-U_0b$ [\cf Eq.~\eqref{eq_def_deltaeff}] depends on the field amplitudes via the bunching parameter $b$ [Eq.~\eqref{eq_ew_cos2}]. This dependence has to be considered in our stability analysis. We assume that
\begin{equation}
  \boldsymbol{\alpha} = \boldsymbol{\alpha}_{\mathrm{ss}} + \boldsymbol{\delta},
\end{equation}
where
\begin{equation}
  \boldsymbol{\delta}=\begin{pmatrix} \delta \\ \delta^{\ast} \end{pmatrix},
\end{equation}
and obtain
\begin{multline}
  \frac{\partial}{\partial t} \left(\boldsymbol{\alpha}_{\mathrm{ss}}+\boldsymbol{\delta}\right) = U(\alpha_{\mathrm{ss}}+\delta,\,\alpha^{\ast}_{\mathrm{ss}}+\delta^{\ast})(\boldsymbol{\alpha}_{\mathrm{ss}}+\boldsymbol{\delta}) + \boldsymbol{\eta}\\
  = U(\alpha_{\mathrm{ss}},\,\alpha^{\ast}_{\mathrm{ss}})\boldsymbol{\delta} + \left[\left.\frac{\partial U(\alpha,\,\alpha^{\ast})}{\partial \boldsymbol{\alpha}}\right|_{\boldsymbol{\alpha}=\boldsymbol{\alpha}_{\mathrm{ss}}}\cdot \boldsymbol{\delta}\right]\boldsymbol{\alpha}_{\mathrm{ss}} + \mathcal{O}(\delta^2),
\end{multline}
where the term in the square parenthesis has to be interpreted componentwise for each element of the matrix $U$. Using $\partial_{\boldsymbol{\alpha}}f=(\partial_nf)\left(\begin{smallmatrix}\alpha^* \\ \alpha\end{smallmatrix}\right)$ with $n=\alpha^*\alpha$, we find
\begin{multline}
  \left[\left.\frac{\partial U(\alpha,\,\alpha^{\ast})}{\partial \boldsymbol{\alpha}}\right|_{\boldsymbol{\alpha}=\boldsymbol{\alpha}_{\mathrm{ss}}}\cdot \boldsymbol{\delta}\right]\boldsymbol{\alpha}_\mathrm{ss}\\=i\left.\frac{\partial \Df}{\partial n}\right|_{n=n_\mathrm{ss}}\left(\alpha^{\ast}_\mathrm{ss}\delta + \alpha_\mathrm{ss}\delta^{\ast} \right)\begin{pmatrix}1 & 0 \\ 0 & -1 \end{pmatrix}\boldsymbol{\alpha}_{\mathrm{ss}} \\
  =i\left.\frac{\partial \Df}{\partial n}\right|_{n=n_\mathrm{ss}} \begin{pmatrix} n_{\mathrm{ss}} & \alpha_{\mathrm{ss}}^2 \\ -[\alpha^{\ast}_{\mathrm{ss}}]^2 & -n_{\mathrm{ss}} \end{pmatrix} \boldsymbol{\delta}.
\end{multline}
Finally, we arrive at
\begin{equation}
  \frac{\partial}{\partial t}\boldsymbol{\delta} = A \boldsymbol{\delta}+\mathcal{O}(\delta^2),
\end{equation}
with the coefficient matrix
\begin{equation}
  A \coloneq U(\alpha_{\mathrm{ss}},\,\alpha^{\ast}_{\mathrm{ss}})+\left.i\frac{\partial \Df}{\partial n}\right|_{n=n_{\mathrm{ss}}} \begin{pmatrix} n_{\mathrm{ss}} & \alpha_{\mathrm{ss}}^2 \\ -[\alpha^{\ast}_{\mathrm{ss}}]^2 & -n_{\mathrm{ss}} \end{pmatrix}.
\end{equation}

\par

Linear stability requires negative real parts of the eigenvalues of $A$. From 
\begin{equation}\label{eq_app_trace}
  \Tr(A) = -2\kappa < 0
\end{equation}
follows, that the eigenvalues of $A$ must be of the form $\lambda_{1,2}=a_{1,2}\pm ib$. At this point we have to discriminate between three cases: (i) $\lambda_{1,2}=a_{1,2}\pm ib$, (ii) complex-conjugate eigenvalues $\lambda_{1,2}=a\pm ib$, and (iii) real eigenvalues $\lambda_{1,2}=a_{1,2}$. The first case can be ruled out since $\det(A)$ is real (see below). In the second case it follows from the negative trace [Eq.~\eqref{eq_app_trace}] that $a<0$, since $\Tr(A)\equiv2a$, i.e., linear stability is always ensured. The third case implies $\det(A)\equiv a_1a_2$, which is positive for negative eigenvalues. We thus require $\det(A)\stackrel{!}{>}0$, which evaluates to
\begin{equation}
  \kappa^2+\left[\Df +\left.n_{\mathrm{ss}} \frac{\partial \Df}{\partial n}\right|_{n=n_{\mathrm{ss}}} \right]^2 - n_{\mathrm{ss}}^2\left( \frac{\partial \Df}{\partial n}\right)^2 \stackrel{!}{>}0.
\end{equation}
Inserting the steady-state photon number [\cf Eq.~\eqref{eq_app_alpha_ss}]
\begin{equation}
  n_{\mathrm{ss}}=\frac{\eta^2}{\kappa^2+\Df^2},
\end{equation}
simple arithmetic yields the condition
\begin{equation}
  \kappa^2+\Df^2+2\Df \frac{\eta^2}{\kappa^2+\Df^2} \left.\frac{\partial \Df}{\partial n} \right|_{n=n_{\mathrm{ss}}} >0,
\end{equation}
which can be recast into the form
\begin{equation}
  1-\frac{\partial}{\partial n}\left.\frac{\eta^2}{\kappa^2+\Df^2}\right|_{n=n_{\mathrm{ss}}} >0.
\end{equation}
This is precisely the inequality~\eqref{eq_selconstab}.

\end{document}